\documentclass{Interspeech2024}

\interspeechcameraready 
\usepackage{graphicx} 
\usepackage{float} 
\usepackage{subfigure} 
\usepackage{diagbox}
\usepackage{algorithm}
\usepackage{algorithmic}
\usepackage{multirow}
\usepackage{makecell}
\usepackage{amsmath}
\usepackage{caption}
\usepackage{cite}
\usepackage{color}
\usepackage{url,array}
\usepackage{amssymb}
\usepackage{hyperref}
\usepackage{booktabs}
\usepackage{amsmath}
\usepackage{hyperref}
\usepackage{bbding}
\usepackage{algorithm}
\usepackage{algorithmic}

\title{NPU-NTU System for Voice Privacy 2024 Challenge}

\name[affiliation={1}]{Jixun}{Yao}
\name[affiliation={2}]{Nikita}{Kuzmin}
\name[affiliation={1}]{Qing}{Wang}
\name[affiliation={1}]{Pengcheng}{Guo}
\name[affiliation={1}]{Ziqian}{Ning}
\name[affiliation={1}]{Dake}{Guo}
\name[affiliation={3}]{Kong Aik}{Lee}
\name[affiliation={2}]{Eng-Siong}{Chng}
\name[affiliation={1}]{Lei}{Xie}

\address{
  $^1$Audio, Speech and Language Processing Group (ASLP@NPU)\\School of Computer Science, Northwestern Polytechnical University\\
  $^2$Nanyang Technological University \\
  $^3$The Hong Kong Polytechnic University}

\email{}

\keywords{speaker anonymization, voice privacy 2024 challenge, voice conversion}

\begin{document}

\maketitle

\begin{abstract}
    
 Speaker anonymization is an effective privacy protection solution that conceals the speaker's identity while preserving the linguistic content and paralinguistic information of the original speech. To establish a fair benchmark and facilitate comparison of speaker anonymization systems, the VoicePrivacy Challenge (VPC) was held in 2020 and 2022, with a new edition planned for 2024. In this paper, we describe our proposed speaker anonymization system for VPC 2024. Our system employs a disentangled neural codec architecture and a serial disentanglement strategy to gradually disentangle the global speaker identity and time-variant linguistic content and paralinguistic information. We introduce multiple distillation methods to disentangle linguistic content, speaker identity, and emotion. These methods include semantic distillation, supervised speaker distillation, and frame-level emotion distillation. Based on these distillations, we anonymize the original speaker identity using a weighted sum of a set of candidate speaker identities and a randomly generated speaker identity. Our system achieves the best trade-off of privacy protection and emotion preservation in VPC 2024.
\end{abstract}

\section{Introduction}

Speech data on the Internet have proliferated exponentially due to the advent of social media, which encapsulates a wealth of sensitive personal information such as the speaker’s identity, age, gender, health status, personality, racial or ethnic origin, geographical background, social identity, and socio-economic status. This sensitive information can be recognized by speaker identification~\cite{zhou2021resnext}, pathological condition detection~\cite{schuller2013interspeech}, or other speech attribute recognition systems. With new regulations like the General Data Protection Regulation (GDPR) in the European Union~\cite{nautsch2019gdpr}, the privacy protection of personal data has gained more attention.

\textit{Speaker anonymization} serves as a proactive privacy protection solution, implemented before users share their speech data. Its fundamental objective is to effectively remove a speaker's identity while preserving the para-linguistic information of the original speech. To establish a fair benchmark and make speaker anonymization systems comparable, the VoicePrivacy Challenge (VPC) was introduced by the speech community and held in 2020 and 2022~\cite{vpc2020eval, vpc2020intro,vpc2022eval}. Previous VPC clearly defines the speaker anonymization task, benchmark, metrics, and datasets to drive the development and innovation of techniques dedicated to preserving the privacy of speech data.

Unlike previous VPC editions, the third edition focuses on preserving the emotional state, a key paralinguistic attribute in many real-world voice anonymization scenarios, such as call centers using third-party speech analytics~\cite{vpc2024}. To evaluate the performance of emotion preservation, the unweighted average recall (UAR) for speech emotion recognition (SER) replaces the previous voice distinctiveness and intonation metrics. Pre-trained models for utility evaluation (linguistic content and emotion) are trained on original data rather than anonymized data to ensure linguistic and emotional content remains undistorted. Additionally, all data are anonymized at the utterance level. The utterance-level anonymization means that the voice anonymization system must assign a pseudo-speaker to each utterance independently of the other utterances~\cite{vpc2024}.

In this paper, we propose a speaker anonymization system based on a disentangled neural codec. We introduce a serial disentanglement strategy to perform step-by-step disentangling from a global time-invariant representation (speaker identity) to a temporal time-variant representation (linguistic content and fundamental frequency). Since the duration of anonymized speech does not change after anonymization, we assume that the emotional characteristics correspond closely to the fundamental frequency. Therefore, we introduce frame-level emotion distillation to disentangle the emotion-related representation, enhancing emotion preservation during the anonymization process. Additionally, we employ a semantic teacher and self-supervised speaker distillation to disentangle linguistic content and speaker identity information. During the anonymization process, the anonymized speaker identity is the weighted sum of a set of candidate speaker identities and a randomly generated speaker identity. Experimental results on VPC 2024 Challenge datasets demonstrate that our proposed system effectively protects speaker identity while maintaining the original linguistic content and paralinguistic information.

\section{Related Works}
The VPC series has shown that most current speaker anonymization systems can be broadly categorized into two classes: (1) signal-processing-based approaches~\cite{sig1,sig2} and (2) neural voice conversion based approaches~\cite{nn4,nn6,nn7,yao2023distinguishable, yao2024distinctive,lv2023salt,yao2024musa}. We provide an overview of these two classes of speaker anonymization approaches in this section.

\begin{figure*}[ht]
  \centering
  \includegraphics[width=17cm]{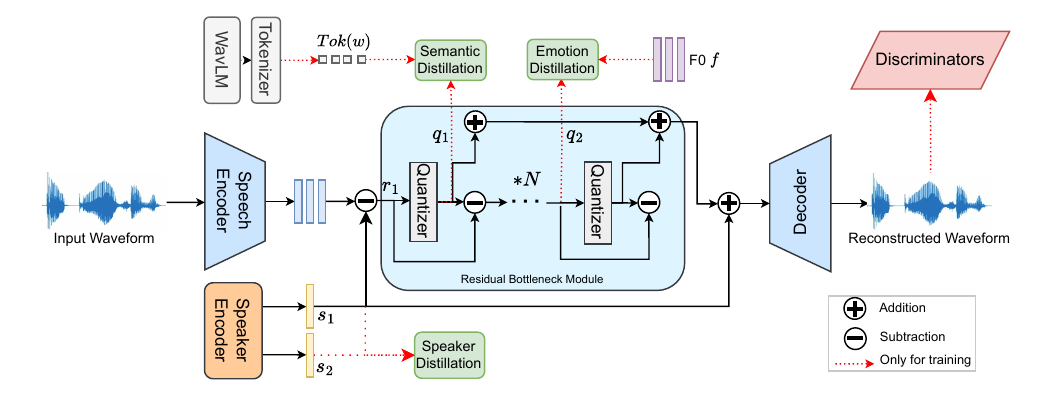}
  \caption{ An overview architecture of our proposed speaker anonymization system. The pre-trained WavLM model is frozen during training and the red dashed lines are only used in the training process.}
  \label{fig:model}
\end{figure*}

\subsection{Signal-processing Based Speaker Anonymization}

Most signal-processing-based approaches do not require training data and directly manipulate instantaneous speech characteristics. The current mainstream method focuses on adjusting formant frequency and F0. A typical approach uses McAdams coefficients to introduce randomized shifts to the formant frequencies~\cite{sig1}, effectively distorting the spectral envelope to anonymize the original speaker's identity. Building on this McAdams coefficients approach, Gupta \textit{et al.} further extends the distortion of the spectral envelope by amplifying the width of formant peaks~\cite{sig2}.

\subsection{Neural Voice Conversion Based Speaker Anonymization}

Neural voice conversion-based speaker anonymization generally outperforms signal-processing-based approaches in terms of privacy and utility metrics. A typical approach involves using a pre-trained ASV model to extract speaker identity representations~\cite{nn4}, like x-vectors~\cite{xvector} or d-vectors~\cite{dvector}, while an ASR model extracts linguistic content representations. The original speaker representation is then replaced by averaging a set of candidate speaker representations from an external pool~\cite{nn6,nn7}. Candidates are selected based on cosine distance to ensure the farthest representations, protecting privacy. Finally, the averaged speaker representation, linguistic content, and F0 are processed by the neural voice conversion model to generate an anonymized mel-spectrogram. A vocoder then converts the anonymized mel-spectrogram into anonymized speech.

\section{Proposed System}

\subsection{Overview}
As illustrated in Figure~\ref{fig:model}, our proposed speaker anonymization system utilizes an auto-encoder architecture, which contains a speech encoder, a speaker encoder, a decoder and a residual bottleneck module. The speech encoder converts the speech samples into frame-level representations including linguistic content, speaker identity, and emotional state. We then serially disentangle each factor, ultimately reconstructing the original waveform from the disentangled components. Specifically, the speaker identity is extracted by the speaker encoder, while the residual bottleneck module disentangles the linguistic content and emotional state. Each disentangled factor is concatenated and used to reconstruct the original speech waveform through the decoder.

During inference, we first disentangle the linguistic content and emotional state from the speech waveform. We then average a set of randomly selected speaker identities and combine the averaged speaker identity with a randomly generated speaker identity to obtain the final anonymized speaker identity. The decoder then combines the original linguistic content and emotional state with anonymized speaker identity to reconstruct the final anonymized speech waveform.

\subsection{Factor Distillation}
In this section, we will introduce the disentanglement details
and three key distillation methods.

\textbf{Speaker distillation}. The speaker identity extraction is a crucial step in achieving speaker anonymization. The speaker’s identity is a voice's inherent and time-invariant characteristic.
We first transform the speech waveform to a mel-spectrogram and feed it into our speaker encoder to produce a global representation $s \in \mathbb{R}^{d}$, while $d$ represents the dimension of the speaker representation. 
To ensure the speaker encoder captures speaker identity information rather than other time-invariant information, we employ two speaker distillation losses to constrain the speaker encoder. We randomly sample two segments from the same utterance and produce two global representations with the speaker encoder. Since the speaker's identity is time-invariant, representations extracted from different segments should be similar and belong to the same speaker identity. We use cosine similarity and an additional speaker classifier with explicit labels to align the extracted global representation with the speaker’s identity. The overall speaker distillation loss $\mathcal{L}_{\textrm{spk}}$ can be defined as follows:
\begin{equation}
    \mathcal{L}_{\textrm{spk}}=\mathbb{E}[-log(C(I\mid s_1))]+\mathbb{E}[-log(C(I\mid s_2))]-\text{cos}(s_1, s_2), 
\end{equation}
where $s_1$ and $s_2$ represent the global representation extracted from two segments, respectively. $I$ is the speaker identity label and $\text{cos}(\cdot)$ denotes cosine similarity. $C(\cdot)$ denote a speaker identity classifier tasked with determining whether $s_*$ is associated with the corresponding speaker identity. Finally, we obtain the speaker identity representation and can further disentangle the speaker identity from the output of the speech encoder by direct subtraction. Thus, we obtain the speaker-independent representation $r_1$.

\textbf{Linguistic distillation}. We utilize a residual bottleneck module for hierarchical disentanglement to further disentangle linguistic content and emotion state from $r_1$. The residual bottleneck module consists of $N$ Vector Quantization (VQ) layers~\cite{van2017neural}, with each VQ layer cascading in a residual manner. Each quantizer contains only residual information from the preceding quantizer, making it inherently suitable for serial disentanglement. Therefore, we distill the first quantizer with a linguistic teacher. We extract the 6th layer's output from the pre-trained WavLM model~\cite{chen2022wavlm} and use a K-means cluster to transfer WavLM's hidden features into discrete tokens, serving as the linguistic teacher. The linguistic distillation loss $\mathcal{L}_{\textrm{lin}}$ can be described as follows:
\begin{equation}
    \mathcal{L}_{\textrm{lin}}=\mathbb{E}[-log(\text{Tok}(w)\mid q_1))],
\end{equation}
where $\text{Tok}(\cdot)$ denote as K-means cluster, $q_1$ and $w$ represent the quantized token from the first quantizer and WavLM's hidden feature, respectively. The linguistic distillation process is inspired by SpeechTokenizer~\cite{zhang2023speechtokenizer}.

\textbf{Emotion distillation}. After aligning the first quantizer to linguistic content, we further constrain the residual quantizer with the fundamental frequency (F0). Since the duration of the anonymized speech does not change after anonymization, we believe the emotional state corresponds closely to the F0. Therefore, we directly use F0 to distill the residual quantizer as follows:
\begin{equation}
    \mathcal{L}_{\textrm{emo}}=cos(f,\text{Proj}(q_2)),
\end{equation}
where $\text{Proj}(\cdot)$ represent the linear projection. $f$ and $q_2$ represent the F0 and quantized embedding of second quantizer.

\subsection{Training Objectives}
The training objective of our proposed anonymization system includes a reconstruction task with adversarial training and three distillation tasks. The reconstruction loss consists of both time and frequency domain losses. The total reconstruction losses are shown as follows:
\begin{align}
    \mathcal{L}_{\textrm{rec}}=\Vert\text{mel}(X)-\text{mel}(\hat{X})\Vert_1  
    +\Vert\text{mel}(X)-\text{mel}(\hat{X})\Vert_2,
\end{align}
where $X$ and $\hat{X}$ represent the original waveform and reconstructed waveform, respectively. Meanwhile, $\text{mel}(\cdot)$ is an 80-bin mel-spectrogram extraction process using a short-time Fourier transform (STFT). For adversarial loss $\mathcal{L}_{\textrm{adv}}$, we follow the same configuration with HiFi-GAN. Similar to the conventional VQ-based models, we employ a straight-through estimator to optimize the commitment loss between the input feature and quantized feature as follows:
\begin{equation}
    \mathcal{L}_{\textrm{com}}=\sum_{i=1}^{N}||x_i-q_i||_2^2,
\end{equation}
where $i$ represents the number of quantizer layers.

Generally, our proposed anonymization system is optimized by the mixture of the following losses:
\begin{align}
\mathcal{L}=\lambda_r\mathcal{L}_{\textrm{rec}}
    +\lambda_a\mathcal{L}_{\textrm{adv}}
    +\lambda_c\mathcal{L}_{\textrm{com}} \notag\\ 
    +\lambda_s\mathcal{L}_{\textrm{spk}}
    +\lambda_l\mathcal{L}_{\textrm{lin}}
    +\lambda_e\mathcal{L}_{\textrm{emo}},
\end{align}
where $\lambda_r$, $\lambda_a$, $\lambda_l$, $\lambda_c$, $\lambda_s$, and $\lambda_e$ are hyper-parameters used to balance each loss term.

\subsection{Anonymization Strategy}
Our proposed anonymization strategy involves combining the averaged speaker identity and a randomly generated speaker identity using weighted sums. To generate the averaged speaker identity, we randomly select a set of speaker identities from the speaker vector pool, following the same configuration as the conventional VPC baseline setting~\cite{vpc2020eval,vpc2022eval}. Additionally, we sample a random speaker identity from a Gaussian distribution. The original speaker identity is anonymized as follows:
\begin{equation}
    s_{\text{anon}} = \alpha\Bar{s}+(1-\alpha)\hat{s},
\end{equation}
where $\alpha$ denotes the weight parameter, $\Bar{s}$ and $\hat{s}$ represent the averaged and randomly sampled speaker identities, respectively.

\begin{table*}[ht]
\centering
\caption{Averaged results over baseline systems and our proposed anonymization system on VPC 2024 development and test datasets. F and M represent the female and male, respectively. Avg denotes average results between female and male EER results and B* denotes different baseline systems described in VPC 2024. C* represents the different EER conditions under VPC 2024 setting and S* represents the different EER conditions under the semi-informed scenario. } \label{tab:eer}
\renewcommand\arraystretch{1.3}
\resizebox{1.0\linewidth}{!}{
\begin{tabular}{lcccccccccc}
\hline
      & \multicolumn{6}{c}{EER, \% ($\uparrow$)}                                              & \multicolumn{2}{c}{WER, \% ($\downarrow$)}                   & \multicolumn{2}{c}{UAR, \% ($\uparrow$)}                                          \\ \cline{2-11} 
      & \multicolumn{3}{c}{LibriSpeech-dev} & \multicolumn{3}{c}{LibriSpeech-test} & \multirow{2}{*}{LibriSpeech-dev} & \multirow{2}{*}{LibriSpeech-test} & \multirow{2}{*}{IEMOCAP-DEV} & \multirow{2}{*}{IEMOCAP-TEST} \\ \cline{2-7}
      & F          & M          & Avg       & F          & M          & Avg        &                                  &                                   &                                  &                                   \\ \hline
Orig. & 10.51       & 0.93      & 5.72      & 8.76        & 0.42       & 4.59        & 1.80                             & 1.85                              & 69.08                            & 71.06                             \\
B1    & 10.94       & 7.45      & 9.20      & 7.47        & 4.68       & 6.07        & 3.07                             & 2.91                              & 42.71                            & 42.78                             \\
B2    & 12.91       & 2.05      & 7.48      & 7.48        & 1.56       & 4.52        & 10.44                            & 9.95                              & 55.61                            & 53.49                             \\
B3    & 28.43       & 22.04     & 25.24     & 27.92       & 26.72      & 27.32        & 4.29                             & 4.35                              & 38.09                            & 37.57                             \\
B4    & 34.37       & 31.06     & 32.71     & 29.37       & 31.16      & 30.26        & 6.15                             & 5.90                              & 41.97                            & 42.78                             \\
B5    & 35.82       & 32.92     & 34.37     & 33.95       & 34.73      & 34.34        & 4.73                             & 4.37                              & 38.08                            & 38.17                             \\
B6    & 25.14       & 20.96     & 23.05     & 21.15       & 21.14      & 21.14        & 9.69                             & 9.09                              & 36.39                            & 36.13                             \\ \hline
C2    & 33.25       & 25.81     & 29.53     & 30.61       & 28.33      & 29.47        & 2.56                             & 2.66                              & 65.98                            & 64.48                             \\
C3    & 38.16       & 33.44     & 35.80     & 32.96       & 30.82      & 31.89        & 3.51                             & 3.19                              & 62.93                            & 60.87                             \\ \hline
S3    & 44.18       & 31.20     & 37.69     & 37.96       & 36.03      & 36.99        & 2.56                             & 2.66                              & 65.98                            & 64.48                             \\
S4    & 45.31       & 39.60     & 42.45     & 40.66       & 40.26      & 40.46        & 3.51                             & 3.19                              & 62.93                            & 60.87                             \\ \hline
\end{tabular}
}
\end{table*}

\section{Experiments Setup}
\subsection{Datasets}
For model training, our anonymization system is trained on both the LibriSpeech~\cite{panayotov2015librispeech} and LibriTTS~\cite{zen2019libritts} datasets. For evaluation, we use LibriSpeech-dev-clean and LibriSpeech-test-clean for privacy and utility evaluation, while the IEMOCAP~\cite{busso2008iemocap} development and evaluation sets are used for emotion evaluation. 

\subsection{Metrics}
The equal error rate (EER) is used as the privacy metric, and the word error rate (WER) and unweighted average recall (UAR) are used as utility metrics. A higher EER indicates better privacy protection, a lower WER indicates better intelligibility, and a higher UAR indicates better emotion preservation. 
Furthermore, we employ two types of privacy protection scenarios: semi-informed and white box. These settings follow VPC 2022~\cite{vpc2022eval} and VPC 2024~\cite{vpc2024} respectively. In VPC 2024, attackers know all hyper-parameters of the anonymization system, which is a challenging scenario.

\subsection{Configuration}
The speech encoder is structured with four convolution blocks, each integrating a residual unit followed by a down-sampling layer. The number of channels doubles during downsampling, and the strides for the four convolution blocks are set as (2, 4, 5, 8). After the convolution blocks, there is a two-layer LSTM for sequence modeling and a concluding 1D convolution layer with a kernel size of 7 and 512 output channels. The architecture of the decoder mirrors the encoder, utilizing transposed convolutions in place of stride convolutions, with the strides in the reverse order of those in the encoder. The speaker encoder and the residual bottleneck module follow the same architecture in \cite{min2021meta} and~\cite{yang2023hifi}. We employ 8 quantizers in the residual bottleneck module.

For model training, we use the AdamW optimizer with parameters $\beta_1$ = 0.8, $\beta_2$ = 0.99, and weight decay $\lambda$ = 1e-5. The learning rate decay followed a schedule with a decay factor of 0.99 per epoch, starting from an initial learning rate of $2 \times 10^{-4}$. The training process comprised a total of 50k steps, utilizing 4 NVIDIA 3090 GPUs with a batch size of 64 utterances. For the hyper-parameters of training loss, we empirical set $\lambda_r=45$, $\lambda_a=1$, $\lambda_c=0.1$, $\lambda_s=1$, $\lambda_m=1$ and $\lambda_e=1$. Regarding the anonymization process, the speaker identity pool is selected from the VCTK dataset. We randomly select 20 speakers and set $\alpha=0.9$ to meet the EER threshold for condition 3, while condition 4 employs $\alpha=0.8$.

\section{Experimental Results}

\subsection{Privacy Protection}
As shown in Table~\ref{tab:eer}, our proposed anonymization system achieves 37.69\% and 36.99\% on dev and test datasets in condition 3, while the EER results for condition 4 are 42.45\% and 40.46\%, respectively.
These EER results outperform all baseline systems, especially when compared with B1 and B2. Additionally, our proposed speaker anonymization system shows a significant advantage in EER results compared to other baseline systems. This indicates that our proposed system can effectively conceal the original speaker's identity and protect personal privacy.

\subsection{Utility Preservation}

Regarding intelligibility utility, the WER results of the original speech are 1.80\% and 1.85\% on the dev and test sets, respectively. The speech anonymized by our proposed system achieves 2.56\% and 2.66\% in condition 3, that only slightly higher than the original speech, demonstrating that our proposed speaker anonymization system can effectively preserve the linguistic content of the original speech. Compared to other baseline systems, our proposed speaker anonymization system also achieves the lowest WER results, especially when compared to B2. The WER results indicate that B2 is undesirable as it significantly degrades intelligibility. Even though the WER results of condition 4 are higher than condition 3, they are still lower than those of other baseline systems.

For emotional evaluation, our proposed system achieves the highest UAR results compared to other baseline systems, with 65.98\% and 64.48\% for the dev and test sets, respectively. In condition 4, the UAR results of our proposed speaker anonymization system are slightly lower than in condition 3 but still higher than those of other baseline systems.

These results show that our proposed speaker anonymization system achieves better utility preservation than baseline systems, effectively preserving both linguistic content and paralinguistic information. Additionally, the utility results for condition 3 are better than those for condition 4, indicating that there is still a trade-off between privacy and utility.

\section{Conclusion}

In this paper, we propose our speaker anonymization system for VPC 2024. The anonymization system is based on the disentangled neural codec and we employ a serial disentanglement strategy to disentangle the linguistic content, speaker identity and emotion state step-by-step. This serial disentanglement strategy disentangles speech representation from a global time-invariant representation to a temporal time-variant representation. To achieve this disentanglement, we propose three specifically designed distillation methods that rely on semantic teachers or explicit labels to guide the disentanglement process. During the anonymization process, the anonymized speaker identity is the weighted sum of a set of candidate speaker identities and a randomly generated speaker identity. Experimental results on VPC 2024 Challenge datasets demonstrate that our proposed system effectively protects speaker identity while maintaining the original linguistic content and paralinguistic information.

\bibliographystyle{IEEEtran}
\bibliography{mybib}

\end{document}